\documentclass{moriond}

\def\be{\begin{equation}}
\def\ee{\end{equation}}
\def\bea{\begin{eqnarray}}
\def\eea{\end{eqnarray}}

\usepackage{microtype}
\usepackage{xspace} 
\usepackage[hypcap=true]{caption}

\usepackage{graphicx}  
\usepackage{color}
\usepackage{colortbl}
\graphicspath{{./figs/}} 

\usepackage{amsmath} 
\usepackage{amssymb}
\usepackage{amsfonts}
\usepackage{upgreek} 

\newcommand{\Photo}{}
\newcommand{\Rk}{{\ensuremath{R_K}}\xspace}
\newcommand{\BKmumu}{\mbox{\ensuremath{B^+ \rightarrow K^+ \mu^+ \mu^-}}\xspace}
\newcommand{\BKee}{\mbox{\ensuremath{B^+ \rightarrow K^+ e^+ e^-}}\xspace}
\def\invfb   {\ensuremath{\rm{~fb}^{-1}}\xspace}
\def\RKvalue {\ensuremath{0.846\,^{+\,0.042}_{-\,0.039}\,^{+\,0.013}_{-\,0.012}}}
\def\significance {{3.1}\xspace}
\newcommand{\BJpsiK}{\mbox{\ensuremath{B^+ \rightarrow K^+ J/\psi}}\xspace}
\newcommand{\Kll}{\ensuremath{K^+ \ell^+ \ell^-}\xspace}
\newcommand{\qsq}{\ensuremath{q^2}\xspace}
\newcommand{\BKll}{\mbox{\ensuremath{B^+ \rightarrow K^+ \ell^+ \ell^-}}\xspace}
\newcommand{\btosll}{\ensuremath{b \to s \ell^+ \ell^-}\xspace}
\newcommand{\BJpsiKll}{\mbox{\ensuremath{B^+ \rightarrow K^+ J/\psi(\ell^+ \ell^-)}}\xspace}
\newcommand{\HbHsmumu}{\mbox{\ensuremath{H_{b} \rightarrow H_{s} \mu^{+} \mu^{-}}}\xspace}
\newcommand{\HbHsee}{\mbox{\ensuremath{H_{b} \rightarrow H_{s} e^{+} e^{-}}}\xspace}
\newcommand{\dif}{\ensuremath{{\rm d}}\xspace}
\newcommand{\BJpsiKee}{\mbox{\ensuremath{B^+ \rightarrow K^+ J/\psi(e^+ e^-)}}\xspace}
\newcommand{\BJpsiKmumu}{\mbox{\ensuremath{B^+ \rightarrow K^+ J/\psi(\mu^+ \mu^-)}}\xspace}
\def\BR         {\mathcal{B}}
       
\newcommand{\aunit}[1]{\ensuremath{\text{\,#1}}}       
\newcommand{\gev}{\aunit{Ge\kern -0.1em V}\xspace}

\begin{document}

\vspace*{4cm}
\title{UPDATED MEASUREMENT OF $R_{K}$ AT LHCb}

\author{ D. LANCIERINI }

\address{Physik Institut, Universit{\"a}t Z{\"u}rich,\\
Winterthurerstrasse 190, 8057, Z{\"u}rich}

\maketitle\abstracts{
In the Standard Model of particle physics, charged leptons of different flavour couple to the electroweak force carriers with the same interaction strength. This property, known as lepton flavour universality, was found to be consistent with experimental evidence in a wide range of particle decays. Lepton flavour universality can be tested by comparing branching fractions in ratios such as \mbox{$\Rk = \mathcal{B}(\BKmumu)/\mathcal{B}(\BKee)$}. This observable is measured using proton-proton collision data recorded with the LHCb detector at CERN's Large Hadron Collider corresponding to an integrated luminosity of 9\invfb.  For a dilepton invariant mass range of $\qsq \in [1.1,6.0] \gev^{2}$, the measured value of $\Rk=\RKvalue$, where the first uncertainty is statistic and the second systematic, is in tension with the Standard Model predicted value at the $\significance\sigma$ level raising evidence for lepton flavour universality violation in \BKll decays.
}

\section{Introduction}

The Standard Model (SM) of particle physics represents our best understanding of the properties and interactions of fundamental particles. It received thorough experimental confirmation since it's formulation in the 1960's. However, it is not a complete theory of fundamental interactions as it is unable to include a description of gravity and leaves some observed phenomena unexplained, such as the matter-antimatter asymmetry or the apparent dark-matter content of the Universe. It also does not incorporate neutrino oscillations and their non-zero masses. Extensions of the SM, referred to as "new physics" (NP), are formulated to explain these phenomena and they predict the existence of new particles and interactions whose signatures can be observed in direct or indirect searches at particle accelerators. Natural units where $\hbar=c=1$ as well as charge conjugation are implied throughout this proceedings.

\subsection{Rare beauty-quark decays}\label{subsec:bdecays}

\begin{figure}
    \centering
    \includegraphics[width=0.8\textwidth]{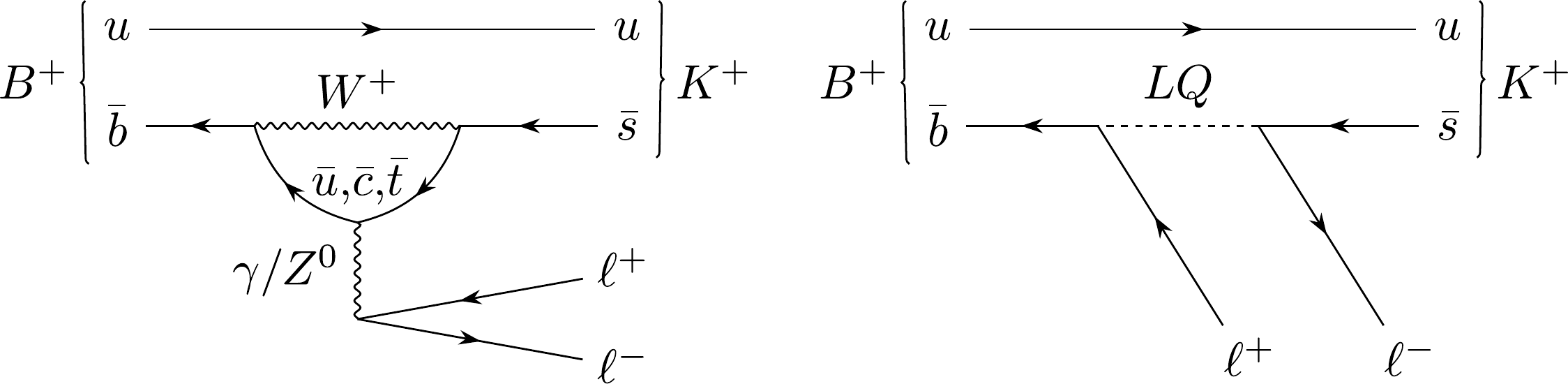}
    \caption[]{Feynman diagrams mediated by left (right) SM (NP) processes. While in the SM contribution to \btosll transitions only occurs at loop level, since it's a flavour-changing neutral current, possible NP models involving a leptoquark \protect\cite{gino} exchange would allow this process to occur at tree level.}
    \label{fig:FDs}
\end{figure}

Signatures of NP can be searched for by comparing decay rates of hadrons, where hadrons are bound states of quarks, with the corresponding SM predictions. Among the wide range of decay processes the SM is able to provide a precise prediction of, transitions where a $B^{+}$ hadron decays into a kaon $K^{+}$, and two charged leptons $\ell^{+}\ell^{-}$ are of great interest. Since at the quark level they involve a \btosll transition, these processes are mediated by flavour-changing neutral currents, which in the SM only occur via electroweak loops (as can be seen in the left-hand side of Fig.~\ref{fig:FDs} and are thus greatly suppressed \cite{GIM}. 
However, virtual contributions from new particles, an example represented on the right hand side of Fig.~\ref{fig:FDs}, could compete with amplitudes of the same strength as the SM, resulting in a modification of the observed branching fraction with respect to the SM prediction.
Rare $B^{+}$ meson decays are thus sensitive to virtual contributions from new particles which could have masses that are inaccessible to direct searches for resonances, even at Large Hadron Collider beam energies.

\subsection{Neutral current anomalies}

Since 2012, several measurements of observables involving \btosll transitions have shown deviations from SM predictions. While there is no single result exhibiting a $5\sigma$ deviation from the SM, the pattern of deviations, collectively denoted as "neutral current anomalies", is striking.
In fact, the latest measurements published by LHCb \cite{isospinKmumu}$^{,}\,$\cite{kstmumu892} all point to a smaller branching fractions for muons in the intermediate \qsq region, where \qsq is the di-lepton invariant mass squared. Precise SM predictions of \btosll processes require good control over non-local QCD effects, such as charmloop contributions, which are known to suffer from potentially large theoretical uncertainties. However, as these non-perturbative effects are lepton flavour universal, they cancel in the ratios of branching fractions $R_{H_{s}}$:
\vspace{-.5em}
\begin{equation}
   R_{H_{s}} = \frac{\int_{\qsq_{\textrm{min}}}^{\qsq_{\textrm{max}}} \frac{\dif \BR(\HbHsmumu)}{\dif\qsq} \dif\qsq}{\int_{\qsq_{\textrm{min}}}^{\qsq_{\textrm{max}}} \frac{\dif \BR(\HbHsee)}{\dif\qsq}\dif\qsq}.
\end{equation}\label{eq:RH}

Where the differential branching fractions are integrated over the di-lepton invariant mass squared range $\qsq \in [\qsq_{\textrm{min}},\qsq_{\textrm{max}}]$ and the final (initial) state $H_{s(b)}$ particles contain a valence strange (beauty) quark.
Due to the universality of the coupling of SM gauge bosons to different lepton flavours, the SM prediction for $R_{H_{s}}$ is unity up to $\mathcal{O}(1\%)$ QED corrections \cite{bip} and negligible phase space effects arising from the different final state lepton masses, hence any sizable deviation from unity would unambiguously indicate breaking of lepton flavour universality (LFU). 

The LHCb collaboration \cite{lhcb1}$^{,}\,$\cite{lhcb2} performed the most precise measurements of LFU sensitive observables, such as $R_{K^{*}}$ \cite{RKstar} and \Rk \cite{prevRK}, and both the results exhibit a tension above $2\sigma$ with respect to the SM expectation. In 2021 LHCb presented an update to the previous measurement of \Rk \cite{2021RK} with the addition of the data collected during 2017 and 2018, effectively doubling size of the analysed data set.

\section{Experimental strategy}\label{sec:expstrat}

The measurement of \Rk is performed using two main ingredients: yields of the decays of interest and the efficiency to select them. However, this involves direct comparison of efficiencies to select muons and electrons decay modes, which interact differently in the detector. While muons traverse the LHCb detector interacting feebly with the material up to the muon stations where they are absorbed, electrons lose a significant amount of energy to bremsstrahlung radiation. This has a degrading effect not only on the mass resolution of the electron with respect to the muon mode, but also on the electrons reconstruction and tracking efficiencies. The main challenge of the measurement consists of controlling for the efficiency of the selection requirements placed to isolate signal candidates and reduce background sources, as these could have systematically different effects on the two lepton flavours. 

In order to suppress such systematic effects, \Rk is measured as a double ratio with respect to the resonant \BJpsiKll mode, which is known to obey LFU to the sub-percent level~\cite{pdg} and has a branching fraction of orders of magnitude larger than the rare \BKll mode:
\vspace{-.2em}
\begin{align}
\begin{split}
    \Rk & =\frac{N(K^{+}\mu\mu)}{N(K^{+}ee)}\frac{\varepsilon(K^{+}ee)}{\varepsilon(K^{+}\mu\mu)} \Big/\underbrace{\frac{N(K^{+}J/\psi(\mu\mu))}{N(K^{+}J/\psi(ee))}\frac{\varepsilon(K^{+}J/\psi(ee))}{\varepsilon(K^{+}J/\psi(\mu\mu))}}_{r_{J/\psi}} \\ 
    & =\frac{N(K^{+}\mu\mu)}{N(K^{+}J/\psi(\mu\mu))}\frac{\varepsilon(K^{+}J/\psi(\mu\mu))}{\varepsilon(K^{+}\mu\mu)} \Big/\frac{N(K^{+}ee)}{N(K^{+}J/\psi(ee))}\frac{\varepsilon(K^{+}J/\psi(ee))}{\varepsilon(K^{+}ee)} \\
\end{split}
\end{align}\label{eq:expRK}

In Eq.~\eqref{eq:expRK}, $N(X)$ and $\varepsilon(X)$ represent respectively the yields and efficiencies of selecting the decay of a $B^{+}$ meson into X. After rearranging the terms in the \Rk definition of Eq.~\eqref{eq:expRK} one can see that the double ratio procedure allows to compare efficiencies to select same lepton flavours in the final states, canceling out systematic uncertainties that are common to both rare and resonant modes thanks to the kinematic overlap of the two decays. The selection requirements are common to both resonant and rare modes up to the cut in $\qsq$ and of the reconstructed mass of the $\Kll$ system. Further selection is applied to reduce background sources and retain only tracks and $B^{+}$ decay vertices with good fit quality.

\subsection{Efficiencies calibration}

Selection efficiencies are estimated using simulated samples, and successively corrected using clean and high statistics calibration data samples. The efficiencies calibration procedure consists of correcting simulation mismodeling using weights that are assigned to simulated events in order to align the selection performance to the one observed in control mode data. The calibration procedure involves correction of the $B^{+}$ kinematics, the efficiency of the particle identification as well as trigger selections and mismatches in the resolution of the \qsq and reconstructed mass of the $B^{+}$ candidate. 

\begin{figure}
\begin{minipage}{0.45\linewidth}
\centerline{\includegraphics[width=\textwidth]{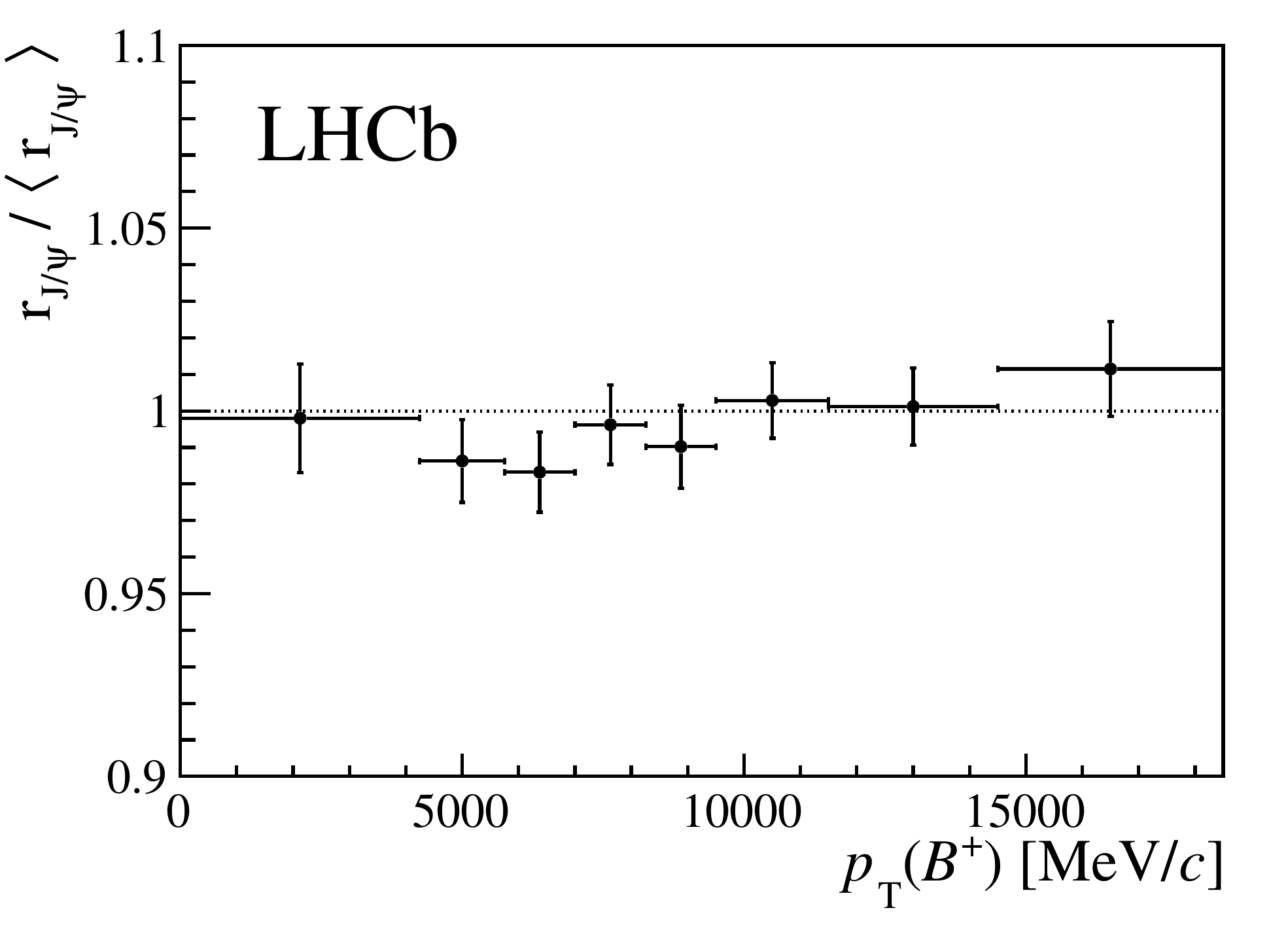}}
\end{minipage}
\hfill
\begin{minipage}{0.45\linewidth}
\centerline{\includegraphics[width=\textwidth]{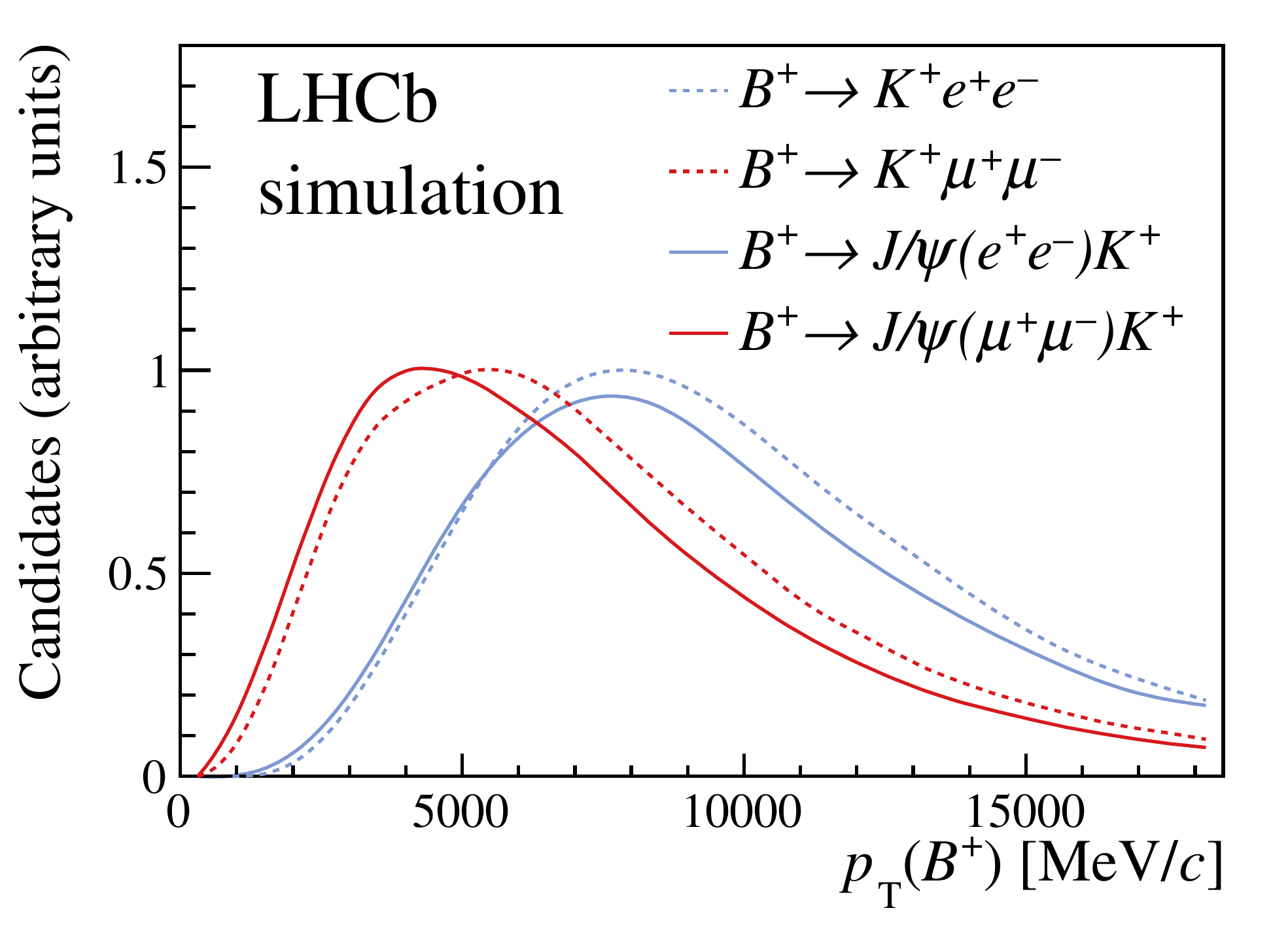}}
\end{minipage}
\caption{The ratio $r_{J/\psi}$ relative to its average value $\langle r_{J/\psi} \rangle$ as a function of $p_{T}$ (left) and distributions of the $B^{+}$ transverse momentum (left). The distribution from the \BJpsiK decays is similar to that of the corresponding \BKll decays such that the differential measurement of $r_{J/\psi}$ tests the kinematic region relevant for the \Rk measurement.}
\label{fig:rjpsidiff}
\end{figure}

\subsection{Procedure validation}

The experimental method is validated through several cross checks that aim at both verifying the control of the selection efficiencies and the cancellation of the systematic effects in the double ratio procedure. Percent-level control of the efficiencies is verified with a direct comparison of the \BJpsiKee and \BJpsiKmumu branching fractions, $r_{J/\psi}$ as defined in Eq.~\eqref{eq:expRK}. This is a stringent cross check of the understanding of the efficiencies as $r_{J/\psi}$ doesn't exploit the systematics uncertainties suppression of the double ratio. The value obtained using the full dataset is found to be $r_{J/\psi} = 0.981 \pm 0.020$ where the error containing both systematic and statistic components is dominated by the systematic uncertainty, as expected. This value is within $1\sigma$ of the LFU expectation.

The control of the efficiencies is further validated by performing $r_{J/\psi}$ single and double-differentially in variables which are relevant to the detector response. An example is shown in Fig.~\ref{fig:rjpsidiff} on the left, where $r_{J/\psi}$ is found to be uniform in bins of transverse momentum of the $B^{+}$ meson. This cross check allows to quantify the effect that the remaining deviations from flatness would have on \Rk if they were due to a genuine mismodelling of the efficiencies rather than statistical fluctuations. Thanks to the large kinematic overlap between the rare and resonant modes, as can be seen in Fig.~\ref{fig:rjpsidiff} (right), the impact on \Rk is within the estimated systematic uncertainty. Similarly, double differential computations of the  $r_{J/\psi}$ ratio do not show any trend and are within the systematic uncertainties assigned to \Rk.

The validity of the efficiencies estimation procedure and the suppression of the systematic effects due to electron muon detection differences is further tested by performing the double ratio:
\vspace{-.5em}
\begin{equation}
R_{\psi(2S)}=\frac{N(K^{+}\psi(2S)(\mu\mu))}{N(K^{+}J/\psi(\mu\mu))}\frac{\varepsilon(K^{+}J/\psi(\mu\mu))}{\varepsilon(K^{+}\psi(2S)(\mu\mu))} \Big/\frac{N(K^{+}\psi(2S)(ee))}{N(K^{+}J/\psi(ee))}\frac{\varepsilon(K^{+}J/\psi(ee))}{\varepsilon(K^{+}\psi(2S)(ee))}
\end{equation}\label{eq:rpsi2S}

The result obtained is $R_{\psi(2S)}=0.997 \pm 0.011$ where the uncertainty includes both statistical and systematic effects. The result is well compatible with the LFU hypothesis for $\psi(2S) \rightarrow \ell^{+}\ell^{-}$ decays and confirms the cancellation of systematic effects to sub-percent level in the double ratio.

\section{Results}

\begin{figure}[!ht]
\begin{minipage}{0.45\linewidth}
\centerline{\includegraphics[width=\textwidth]{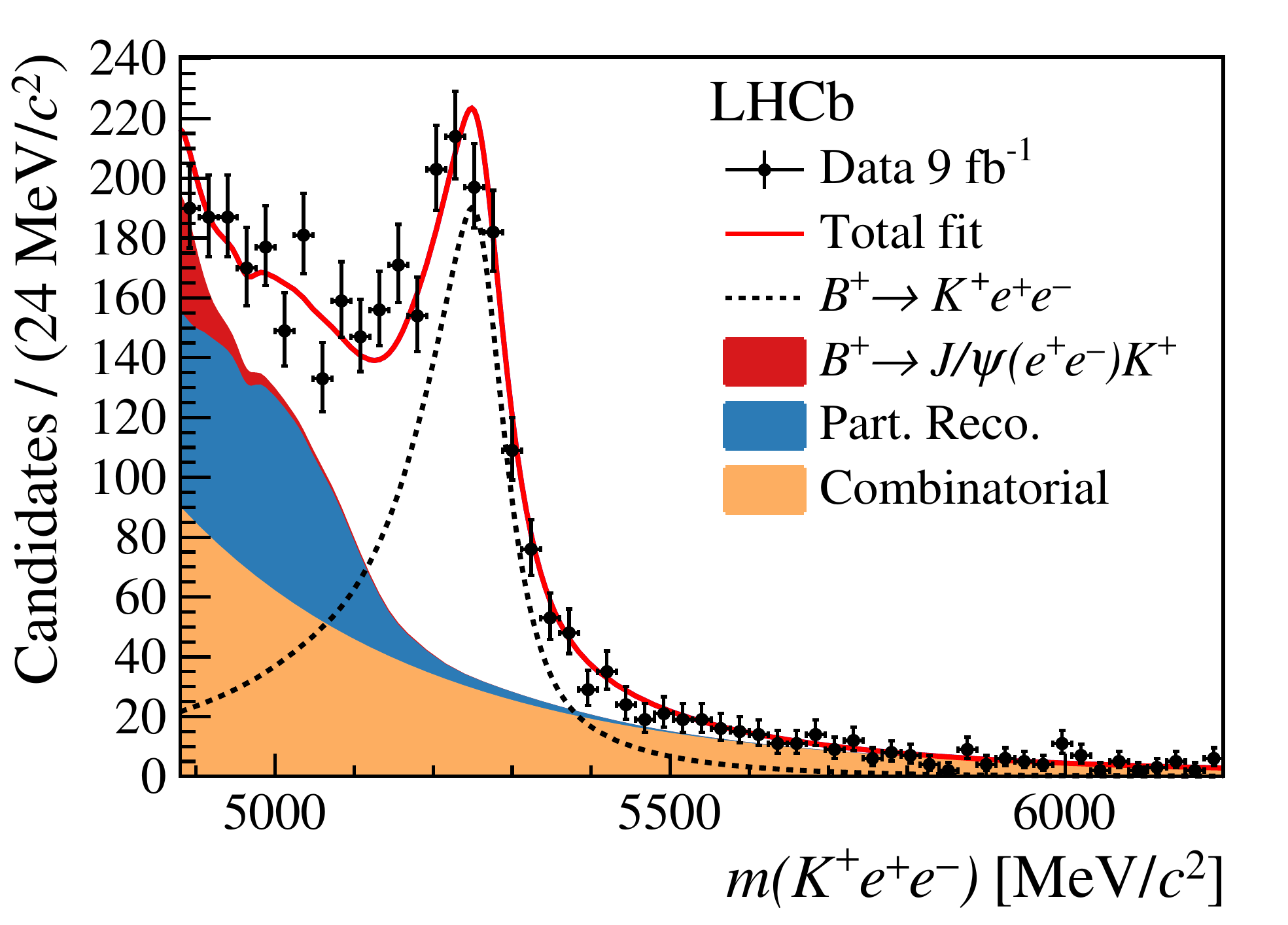}}
\end{minipage}
\hfill
\begin{minipage}{0.45\linewidth}
\centerline{\includegraphics[width=\textwidth]{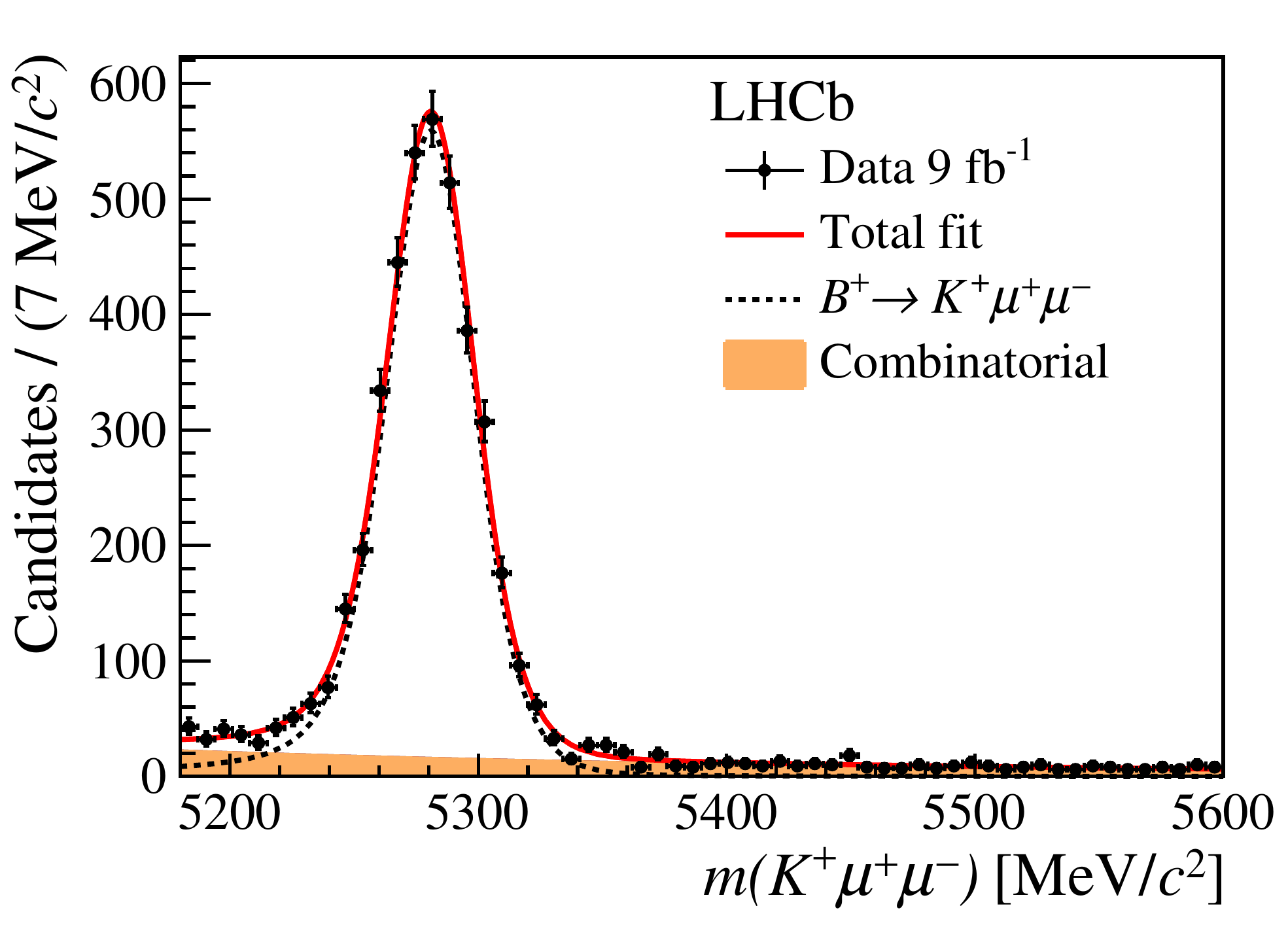}}
\end{minipage}
\caption[]{Distribution of the invariant mass for candidates with (left) electron and (right) muon pairs in the final state for the rare \BKll signal channels.}
\label{fig:rarefit}
\end{figure}

The value of \Rk is extracted as a parameter of interest from an unbinned extended maximum likelihood fit which is performed simultaneously to the \BKmumu and \BKee data, the projections are shown in Fig.~\ref{fig:rarefit}

The mass-shape parameters are derived from the calibrated simulation. It can be noted that the statistics of the muon data sample is higher than the electron case due to lower electron reconstruction efficiency, as discussed in Sec.~\ref{sec:expstrat}. In both muon and electron fits the signal shape forms a peaking structure which in the electron case has a long elongated tail at lower reconstructed $B^{+}$ masses owing to the fact that the bremsstrahlung recovery algorithm is not perfect and not always the energy lost is correctly recovered. Moreover, due to the poorer mass resolution of the electrons, the selected data sample is polluted by backgrounds coming from the resonant \BJpsiKee mode or from partially reconstructed backgrounds such as, for example, $B^{0} \rightarrow K^{*0}(K^{+}\pi^{-})e^{+}e^{-}$ decays in which the $\pi^{-}$ is not reconstructed. In both electrons and muon fits, candidates formed from random track combinations, extend thoughout the fit mass window and correspond to the combinatorial background.

The obtained value of \Rk is corrected by taking into account the bias introduced by the choice of fit model, which is found to be small, and the result reads:

\begin{equation}
    \Rk = \RKvalue
\end{equation}

Where the first uncertainty is statistical, and the second is systematic. As expected the error on the result is dominated by the statistical contribution rather than the systematic one, the dominant contributions to the systematic uncertainties being the choice of the fit model for signal and partially reconstructed decays while the effects due to trigger and kinematics corrections amount to the per mille level showing robustness of the double ratio method.

\begin{figure}
\begin{minipage}{0.45\linewidth}
\centerline{\includegraphics[width=\textwidth]{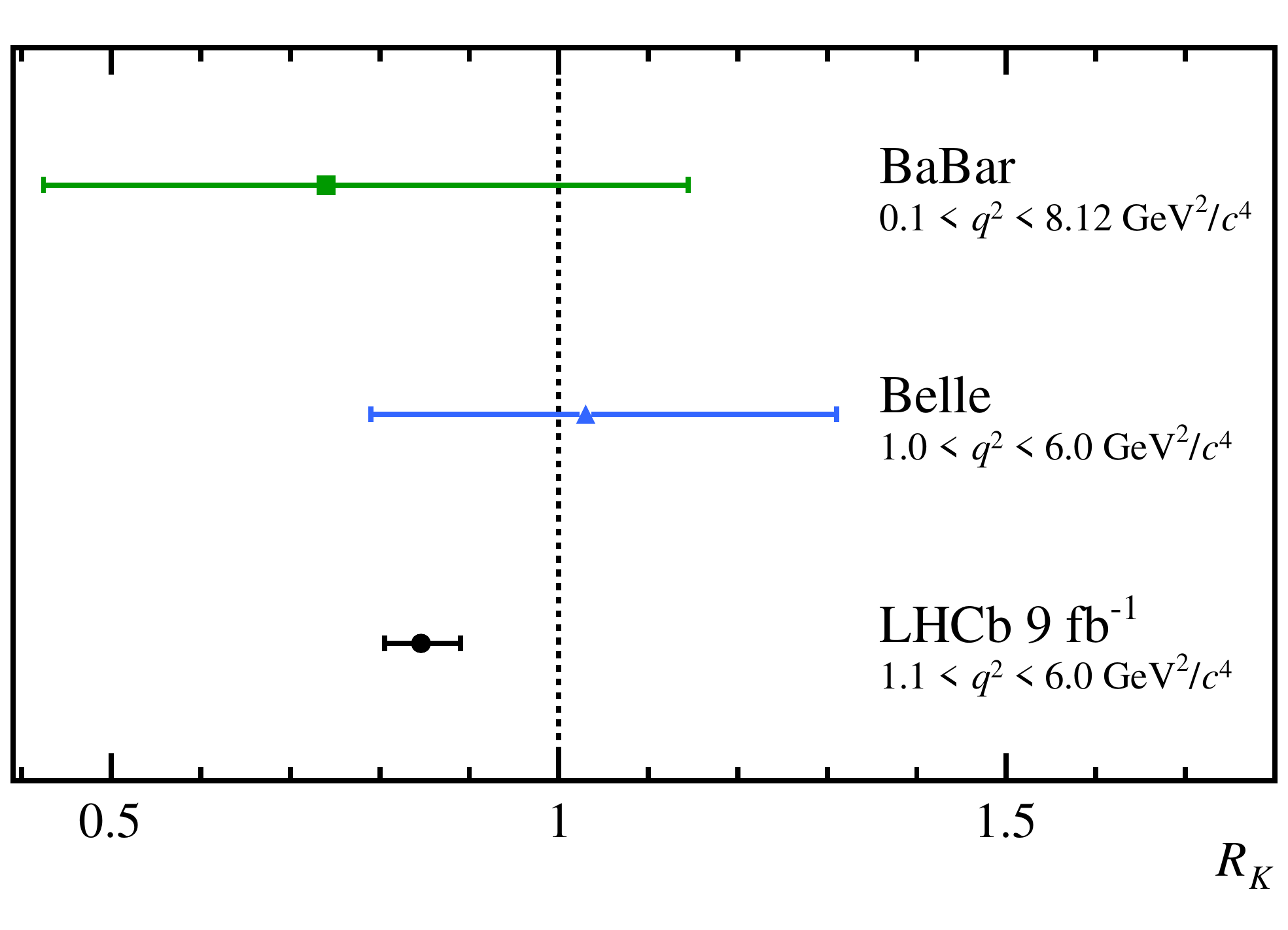}}
\end{minipage}
\hfill
\begin{minipage}{0.45\linewidth}
\centerline{\includegraphics[width=\textwidth]{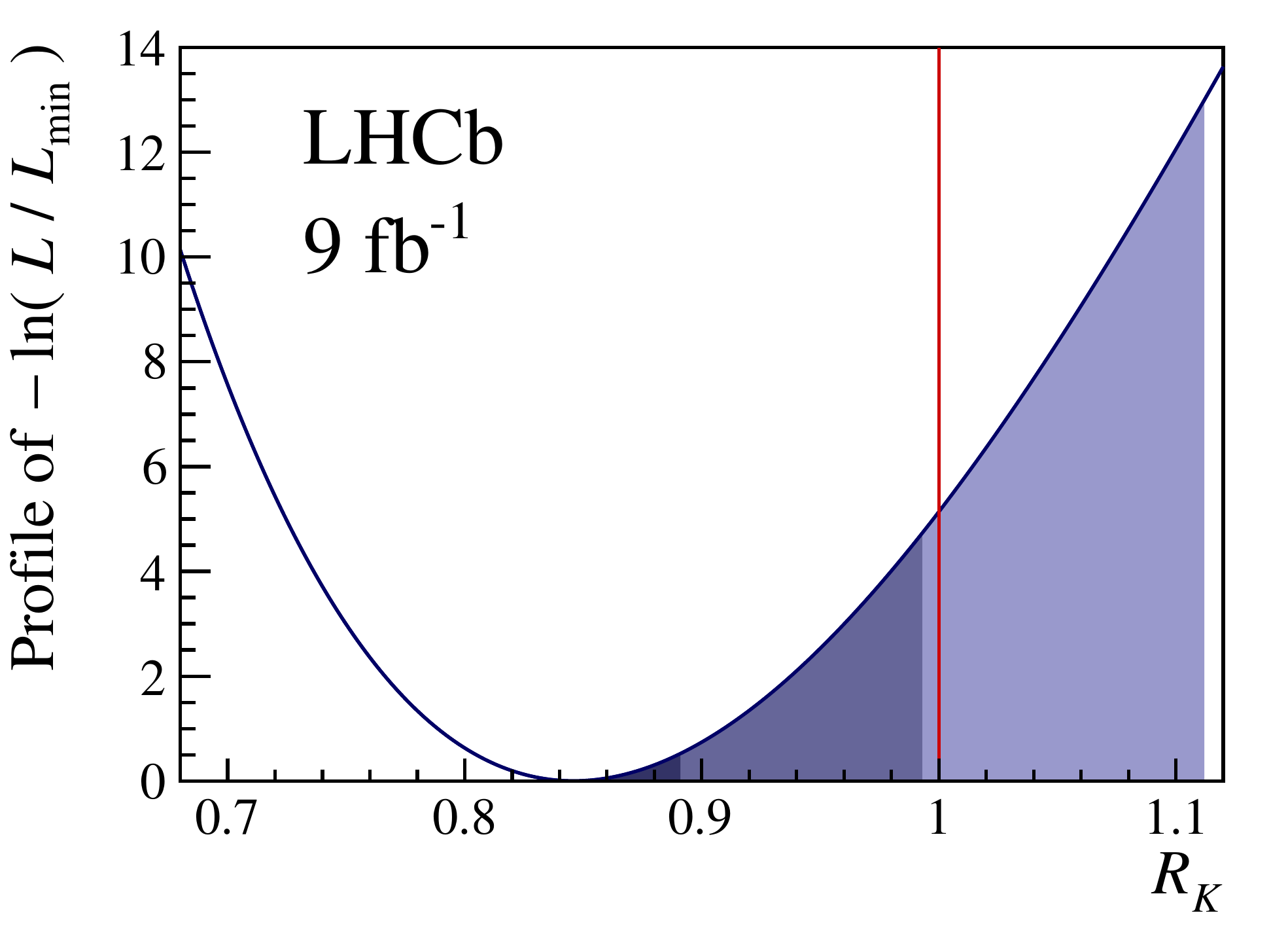}}
\end{minipage}
\caption[]{Left: Comparison between \Rk measurements. In addition to the LHCb result (black), the Babar~\protect\cite{babar} (green) and Belle~\protect\cite{belle} (blue) collaborations are shown. Right: likelihood function from the fit to \BKll candidates profiled as a function of \Rk. The extent of the dark, medium and light blue regions shows the values allowed for \Rk at $1,3,5\sigma$ levels. The red line indicates the SM prediction.}
\label{fig:figresult}
\end{figure}

The obtained result, being the most precise measurement of \Rk to date (Fig.~\ref{fig:figresult}), has a p-value under the SM hypothesis of $0.1\%$. 
This corresponts to a deviation from the SM prediction of $\significance\sigma$, and therefore constitutes the first evidence of LFU violation in \BKll decays obtained with a single experiment.

Another result can be obtained by combining the extracted value of \Rk with the result for the branching fraction measurement $\dif \BR(\BKmumu)/ \dif \qsq$ from \cite{isospinKmumu} to obtain $\dif \BR(\BKee)/ \dif \qsq$ in the same dilepton invariant mass region

\begin{equation}
    \frac{d\mathcal{B}(B^{+}\rightarrow K^{+}e^{+}e^{-})}{\dif q^{2}}= (28.6^{+1.5}_{-1.4} \textrm{ (stat.) } \pm 1.3 \textrm{ (syst.) }) \times 10^{-9}\, \gev^{-2}
\end{equation}

\vspace{-.5em}
\section{Summary}

The observable \Rk is measured using 9 \invfb of proton-proton collision data recorded by the LHCb experiment and the result is found to be in tension with the SM prediction at the level of $3.1\sigma$. This result constitutes evidence for the violation of LFU in \BKll decays.


\begin{thebibliography}{}

\end{thebibliography}


\section*{References}

\begin{thebibliography}{99}
\bibitem{GIM} S. L. Glashow, J. Iliopoulos, and L. Maiani. Weak interactions with lepton-hadron symmetry. \textit{PRD}, 2:1285–1292, Oct 1970.
\bibitem{gino} C. Cornella, D. A. Faroughy, J. Fuentes-Martin, G. Isidori, M. Neubert, Reading the footprints of the B-meson flavor anomalies, arxiv.org:2103.16558.
\bibitem{AltStang} New Physics in Rare B Decays after Moriond 2021 arxiv:2103.13370.
\bibitem{isospinKmumu} R. Aaij et al. Differential branching fractions and isospin asymmetries of $B \rightarrow K^{(*)} \mu^{+}\mu^{-}$ decays. \textit{JHEP} 2014, 133 (2014).
\bibitem{kstmumu892} Measurements of the S-wave fraction in \mbox{$B^{0} \rightarrow K^{+}\pi^{-} \mu^{+}\mu^{-}$} decays and the \mbox{$B^{0} \rightarrow K^{*}(892) \mu^{+}\mu^{-}$} differential branching fraction. \textit{JHEP} 2016, 47 (2016).
\bibitem{lhcb1} A. A. Alves Jr. et al. The LHCb detector at the \textit{LHC. JINST}, 3:S08005, 2008.
\bibitem{lhcb2} R. Aaij et al. LHCb detector performance. \textit{Int. J. Mod. Phys.}, A30:1530022, 2015.
\bibitem{bip} On the standard model predictions for \Rk and $R_{K^{*}}$. \textit{EPJC} 76, 440 (2016).
\bibitem{RKstar} R. Aaij et al. Test of lepton universality with $B^{0} \rightarrow K^{*0} \ell^{+} \ell^{-}$ decays \textit{JHEP} 08 (2017) 055.
\bibitem{prevRK} R. Aaij et al. Search for Lepton-Universality Violation in \BKll Decays. \textit{PRL} 122 (2019) 19, 191801.
\bibitem{2021RK} R. Aaij et al. Test of lepton universality in beauty-quark decays. arXiv:2103.11769.
\bibitem{pdg} P. A. Zyla et al. Review of particle physics. \textit{PTEP}, 2020(8):083C01, 2020.
\bibitem{babar}  J. P. Lees et al.  Measurement of branching fractions and rate asymmetries in the rare decays \mbox{$B \rightarrow K^{(*)} \ell^{+}\ell^{-}$}. \textit{PRD}, 86:032012, 2012.
\bibitem{belle}  S. Choudhury et al. Test of lepton flavor universality and search for lepton flavor violation in $B \rightarrow K \ell \ell$ decays. \textit{JHEP}, 03:105, 2021.


\end{thebibliography}
\end{document}